\documentclass[aps,prd,eqsecnum,twocolumn,twoside,showpacs]{revtex4}
\usepackage{amsfonts}
\begin{document}

\title{ A search for the minimal unified field theory.
II.  Spinor matter and gravity. }
\author{ Alexander Makhlin}
\email[]{amakhlin@comcast.net}
\affiliation{ Rapid Research Co., Southfield, MI, USA}
\date{September 9, 2004}
\begin{abstract}
The spin connections of the Dirac field have three ingredients that
are connected with the Ricci rotations, the Maxwell field, and an
{~\em axial field~} which is coupled to the axial current. I
demonstrate that the  axial field provides an  effective mechanism of
auto-localization of  the Dirac field into compact objects. A
non-linear system of Dirac and sine-Gordon equations that has a
potential to yield a mass spectrum is derived. The condition that the
compact objects are stable (the energy-momentum is self-adjoint)
leads to the Einstein's field equations. The Dirac field with its
spin connection seem to be a natural material carrier of the
space-time continuum in which the compact objects are moving along
the geodesic lines. The long distance effect of the axial field is
indistinguishable from the Newton's gravity which reveals the
microscopic nature of gravity and the origin of the gravitational
mass.
\end{abstract}
\pacs{03.30.+p, 03.65.-w, 03.70.+k, 04.40.-b,12.10.-g,98.80.-k}
\maketitle

\section{ Introduction \label{sec:Sec1}}
\renewcommand{\theequation}{I.\arabic{equation}}
\setcounter{equation}{0}

In this paper, I continue to explore an hypothesis that the Dirac
spinor field is the primary form of matter and that its dynamics is
driven exclusively by its spin connections. In the previous paper
\cite{paper1},  it was shown that a residual degree of freedom, which
is left to the Dirac spinor field by the rule of the Lorentz
transformation of its probability current, allows for the existence
of {\em two} real vector fields. One of these fields, $A^a(x)$, has
the properties of the Maxwell field derived within the standard
scheme of gauge invariance \cite{Fock1,Schrod1,Pauli1}. This field is
minimally coupled to the conserved probability current $j^a$. The
second field, the axial field $\aleph^a(x)$, is minimally coupled to
the axial current $J_5^a\equiv{\cal J}^a$ which, in turn, has a
pseudoscalar density $\cal P$ as a source. This current is not
conserved and $\aleph^a$ is a massive (not gradient invariant) field.
The vector field $A^a$ affects the parallel transport of a
four-component Dirac spinor {\em as a whole}, while the $\aleph^a$
field acts on its left and right components differently. Therefore,
the axial field becomes most visible in the {\em polarization}
characteristics of Dirac spinors, i.e. bilinear forms with the left-
and right- spinor components mixed.

In the context of this study, both fields, $A^a$ and $\aleph^a$, are
derived and treated as the spin connections of the Dirac field
$\psi$. The spinors themselves represent a local (and only local)
group of the Lorentz transformations and the probability current of
two-component spinors is light-like. Therefore, the spinor fields do
not set any limit on the space-time resolution in all conceivable
measurements.  The dynamics of spinors dominate all physical
phenomena where sharp space-time localization occurs. Since the
light-like propagation is an intrinsic property of spinor fields they
provide a material footing for the first axiom of special relativity
-- all signals must travel at the speed of light. The spin
connections naturally inherit this property.

Special relativity also requires (the second axiom) the existence of
inertial observers that have finite sizes and lifetimes so that the
notion of their proper time is sensible. In scattering theory, it is
closely related to the so-called {\em cluster decomposition
principle}. Construction of such finite-sized objects is a
long-standing problem of the relativistic field theory. In this
paper, I show that a modest and natural requirement that  particles
are defined as some configurations of the Dirac field that do not
change for a period of time (are parallel-transported) makes the spin
connection $\aleph^a$ the {\em gradient of a scalar function}. Then
$\aleph^a$ becomes a force that leads to tightly bound configurations
of the spinor field. This effect is so pronounced that one may think
of auto-localization as one of the genuine properties of the Dirac
field. In other words, the Dirac field, with its spin connections,
provides a material support to the whole physical structure, which is
called {\em the space-time continuum}. The gravitational interaction
is initiated by the axial current and is transmitted by the (static
or propagating) axial field and axial polarization; at large
distances, it is indistinguishable from Newton's gravity.

The objective of this paper is to show that a theory based on the
interaction of only three fields, $\psi$, $A^\mu$ and $\aleph^\mu$
is self-consistent. The main result of this work is contained in
Eqs.~(\ref{eq:E3.12}) and (\ref{eq:E3.15}), given below for
convenience,
\begin{eqnarray}
\alpha^a\{\partial_a +ieA_a(x)
+(i/2)\rho_3 \partial_a\Upsilon(x)~~~~\nonumber\\
 -\Omega_a(x)\}\psi +im\rho_1\psi=0,
\label{eq:I.1}\end{eqnarray}
\begin{eqnarray}
\Box\Upsilon(x) -
{4g^2m\over M^2}~{\cal R}(x)\sin\Upsilon(x)=0~,~~\nonumber\\
{\cal R}^2= (\psi^+\alpha^a\psi)^2~.
\label{eq:I.2}\end{eqnarray}
This non-linear system of equations is capable of describing the
interaction and spectrum of masses/scales/ energies of localized
states at short distances as well as propagation of these states as
compact objects. The first of these equations is the linear Dirac
equation which is modified by the presence of an axial field. It can
be considered as a microscopic equation that preserves the
superposition principle. The second equation interpolates between the
short and long distances in a sense that it describes the axial field
subjected to the condition of parallel transport of a compact spinor
object. At large distances, the field $\Upsilon(x)$ becomes the
Newton potential; it is responsible for a residual (gravitational)
interaction between Dirac clusters.

An additional condition that the spinor clusters remain stable
(the Dirac operator and the operator of energy-momentum are
self-adjoint) yields Eqs.~(\ref{eq:E4.7}) and (\ref{eq:E4.4}),
\begin{eqnarray}
R_{\lambda\sigma}=0~,~~~~~~~~~~~~~\label{eq:I.3}\\
\Gamma^\sigma_{\mu\nu}T^\nu_{~\sigma}
=- m {\cal P}~\partial_\mu \Upsilon ~,~\label{eq:I.4}
\end{eqnarray}
where $R_{\lambda\sigma}$ is the Ricci curvature tensor. The first of
these is the Einstein equation for the metric of free space. The
second is a manifestation of local equivalence between the
axial-Newton field and the forces of  inertia. Eq.~(\ref{eq:I.4}) is
the bridge between the physical dynamics of spinor fields and the
origin of dynamical metrics of  general relativity.

There appears to be no problem of {\em including} the spinor fields
{\em into} general relativity. The equations of general relativity
emerge as one of the constituents of spinor dynamics in the limit
when the spinor fields and their spin connections form particles that
may serve as rods and clocks, i.e., the inertial frames which, in
their turn, are necessary to verify the initial hypothesis of local
Lorentz invariance. In agreement with the Einstein's eventual
judgement \cite{EI}, the field equation (\ref{eq:I.3}) has no
energy-momentum tensor of matter in its right-hand side. The
pseudoscalar $\cal P$, which is the maximum of the probability
density localized in a compact spinor object, takes the role of
gravitational mass. {\em The physical origin of the macroscopic
forces of gravity between any two bodies is a trend of the global
Dirac field to concentrate around the microscopic domains where this
field happened to be extremely localized.} These forces tend to
polarize  matter at the level of its spinor organization and may well
play a role at various stages of matter evolution.

The paper is organized as follows: Sec.~\ref{sec:Sec2} is a review of
the results of the previous work \cite{paper1}, which are modified by
incorporating an explicit requirement that the Dirac operator be
self-adjoint. In Sec.~\ref{sec:Sec3}, I formulate a criterion that
the Dirac field represents a compact object. The ansatz of parallel
transport of Dirac particles is introduced, and the form of the spin
connection, compatible with this ansatz, is found. In
Sec.~\ref{sec:Sec4}, I show that the integrity of a Dirac particle
implies that it moves along a geodesic world line of the metric
background that satisfies Einstein's field equation. A model that
demonstrates the auto-localization in a simple spherical geometry is
also worked out in Sec.~\ref{sec:Sec5}.

\section{ Parallel transport of the spinor field \label{sec:Sec2}}

\renewcommand{\theequation}{2.\arabic{equation}}
\setcounter{equation}{0}

 The affine connection in the covariant derivative of a vector field
can be derived in a relatively simple way because rotation of a
vector at a given point follows the rotation of the {\em local}
coordinate axes. Since the Lorentz spinors are defined locally and
only locally and their components are not given in terms of tensor
variables, there is no similar rule for spinor fields. One has to
resort to the so-called tetrad formalism \cite{Eisenhart}, which
conforms to the principle of the equivalence of local inertial frames
in special relativity.

The curvilinear coordinates, which are used throughout this study,
are not connected {\em a priori} with the true curvature of
space-time. The concept of a compact spinor object naturally leads to
the systems of coordinates in which at least one ``radial'' direction
is parameterized by closed two-dimensional surfaces. Even in the
absence of a true space-time curvature, the affine and spin
connections remain indispensable  attributes of this scheme. For a
stable compact object, the transport along closed surfaces can have a
group property.

\subsection{ Spin connection}

The most important physical quantity that provides access to the
geometric properties of spinors is the probability current. Its time
component is a unit operator that commutes with everything. Thus,
it corresponds to the most fundamental measurement of quantum
mechanics. This current, $~j_a =\psi^+ \alpha_a \psi \equiv
\psi^+ (1,\rho_3\sigma_i)\psi \equiv {\bar \psi} \gamma_a \psi~$,
must be a Lorentz vector, which is transformed as
$j_a(x)\to\Lambda_{a}^{~b}(x)j_b(x)$ under a local Lorentz
rotation, and its variation under the parallel displacement
$dx^\mu$ is $~\delta j_\mu=\Gamma^\nu_{\mu\lambda}j_\nu dx^\lambda$.
The tetrad components of this vector change by
$\delta j_a=\omega_{acb}j^c ds^b=\omega_{acb}\psi^+\alpha^c \psi ds^b~$
when this vector is transported by $ds^a$. In these equations,
$\Gamma^\nu_{\mu\lambda}$ are the Christoffel symbols,
$\partial_a=e_a^\mu\partial_\mu$ is the derivative in direction $a$,
and $\omega_{abc}$ are the Ricci rotation coefficients,
($\omega_{bca}=(\nabla_\mu
e_{b}^{\nu})e_{c\nu}e_a^\mu=-\omega_{cba}$).

The convention for Dirac matrices is as follows: The basic matrices
$~\rho_i~$ and $~\sigma_i$, $~i=1,2,3$, were introduced by Dirac
\cite{Dirac1}; we use $~\rho_{0}=\sigma_0={\bf 1}~$, for unit matrix.
The other notations  are: $\alpha_a =(\alpha_0,\alpha_i)$ (with
$\alpha_0=1$, $~\alpha_i=\rho_{3} \sigma_i$),
$\rho_1=\beta=\gamma^0$, $\rho_2=-i\gamma^0\gamma^5$, and
$\rho_3=-\gamma^5$. The $4\times 4$ matrices $\sigma$ and $\rho$
satisfy the same commutation relations as the Pauli matrices, and
all matrices $\sigma$ commute with all matrices $\rho$, i.e.,
$ \sigma_i\sigma_k=\delta_{ik}+i\epsilon_{ikl}\sigma_k$,
 $\rho_a\rho_b=\delta_{ab}+i\epsilon_{abc}\rho_c$, and
$\sigma_i\rho_a -\rho_a\sigma_i =0$.

Let matrix $\Gamma_a$ (the spin connection) define the change of the
spinor components in the course of the same infinitesimal displacement,
$\delta \psi= \Gamma_a \psi ds^a$, $~\delta \psi^+=\psi^+ \Gamma^+_a ds^a$.
This gives yet another expression for $\delta j_a$, namely,
$\delta j_a= \psi^+ (\Gamma^+_b \alpha_a+\alpha_a\Gamma_b)\psi ds^b$.
The two forms of $\delta j_a$ must be the same. Hence,
the equation that defines $\Gamma_a$ is
\begin{eqnarray}
\Gamma^+_b \alpha_a+\alpha_a\Gamma_b= \omega_{acb} \alpha^c ~,
\label{eq:E2.0}\end{eqnarray}
and it has been shown to have the most general solution \cite{paper1},
\begin{eqnarray}
\Gamma_b(x)=-ieA_b(x)-ig\rho_3 \aleph_b(x)+ \Omega_b(x)~,
\label{eq:E2.1}\end{eqnarray}
where the last term is the geometric part,
\begin{eqnarray}
\Omega_b(x)=
{1\over 4}~\omega_{cdb}(x)\rho_1\alpha^c\rho_1\alpha^d, \nonumber
\end{eqnarray}
and the signs of coupling constants are chosen with the electron in mind.
The covariant derivative of a spinor now reads as
$$D_a\psi=(\partial_a-\Gamma_a)\psi,~~~\Gamma_\mu=e^a_\mu\Gamma_a,$$
$$D_\mu\psi=e^a_\mu D_a\psi=(\partial_\mu-\Gamma_\mu)\psi,~$$
in tetrad- and coordinate-basis, respectively.

It is useful to remember that the absolute differential, $DV_a \equiv
D_cV_a d s^c$, of a vector $V_a$ is the principal linear part of the
vector increment with respect to its change in the course of a
parallel transport along the same infinitesimal path. Therefore, the
parallel transport just means that $DV_a=0$; the vector does not
change. The absolute differential of a Dirac field is needed for
exactly the same reason. A stable spinor object does not change and
it {\em is} parallel-transported along its world line. (As the matter
of fact, the notion of parallel transport requires only one
particular curve and a connection on this curve. The full congruence
of curves in the vicinity of the path of parallel transport is not
needed.) The tetrad representation most adequately reflects the local
nature of vector fields in relativistic field theory. The law of the
Lorentz transformation and the definition of parallel transport for
spinors are far less obvious, because their components are not
directly connected with the vectors of the coordinate axes.
Therefore, spinors should always be treated as coordinate scalars.
All Dirac matrices are treated as pure number constructs, which are
the tools for certain substitutes of spinor components.

Although I called $A_a(x)$ and $\aleph_a(x)$ the fields, it seems
more physical to give these designations to
$e(\rho_0)_{\alpha\beta}A_a=\delta_{\alpha\beta}eA_a$ and
$g(\rho_3)_{\alpha\beta} \aleph_a$,  the tetrad components of the
spin connections. This is in line with the fact that the fields are
always measured by their action on material bodies and that this
action is always detected through the kinematics of the material
bodies' motion.  From this perspective, one may  conclude that the
physical meaning are not even the fields $\delta_{\alpha\beta}eA_a$
and $g(\rho_3)_{\alpha\beta}\aleph_a$ but rather the matrix elements
of the full spin connection $(\Gamma^a)_{\alpha\beta}$ between
various states of the spinor field.

\subsection{ Parallel transport of bilinear forms.}

A summary of the computation of covariant derivatives for the
sixteen basic bilinear spinor forms is as follows. If
${\cal J}_a=\psi^+ \rho_3\alpha_a \psi
={\bar \psi}\gamma^5 \gamma_a \psi$  is the axial current,
${\cal S}={\bar \psi}\psi = \psi^+ \rho_1 \psi$
and $~{\cal P}={\bar \psi}\gamma^5\psi = \psi^+ \rho_2 \psi$
are the two Lorentz scalars, and $M_{ab}$ is a tensor with
components $M_{0i}= -i\psi^+ \rho_2\sigma_i \psi
= {\bar \psi}\gamma^0\gamma^i\psi$ and
$M_{ik}= -i\epsilon_{ikm}\psi^+ \rho_1\sigma_m \psi
 ={\bar \psi}\gamma^i\gamma^k\psi$, then
\begin{eqnarray}
D_\mu j^\nu=\nabla_\mu j^\nu~,~~~~
D_\mu {\cal J}^\nu=\nabla_\mu {\cal J}^\nu~, \nonumber \\
D_\mu M^{\lambda\nu}=\nabla_\mu M^{\lambda\nu}
+g~ \aleph_\mu
\epsilon^{\lambda\nu\rho\sigma}M_{\rho\sigma}, \nonumber \\
D_\mu {\cal S}=\nabla_\mu {\cal S}+2g~\aleph_\mu {\cal P},~~
D_\mu {\cal P}=\nabla_\mu {\cal P}-
2g\aleph_\mu {\cal S}~.
\label{eq:E2.2}\end{eqnarray}
These equations are originally derived in the tetrad representation
(when only spinors are differentiated, $D\psi=D_a\psi ds^a$) and
only after that are translated into coordinate form.
The first equation  duplicates the input for
Eq.~(\ref{eq:E2.0}). The parallel transport of the vector
currents that are built with the aid of diagonal Dirac matrices
$\sigma_i$ and $\rho_3$ is not affected by the axial field.
In the Lorentz scalars and the tensor, the left- and right- spinors are
mixed by either $\rho_1$ or $\rho_2$ which makes their covariant
derivative dependent of $\aleph_\mu=e_\mu^a \rho_3 \aleph_a$.

To derive equations similar to (\ref{eq:E2.2}) for any other form
$O$, defined by an operator ${\cal O}$, one has to follow a simple rule,
\begin{eqnarray}
DO(x) \equiv  D [\psi^+ {\cal O} \psi] =
\psi^+ ({\cal O}\overrightarrow {D_a} +
\overleftarrow{D^+_a}{\cal O}) \psi~ds^a~.
\label{eq:E2.3}\end{eqnarray}
Accordingly, the parallel transport of $O(x)$ means that $DO(x)=0$.

The vector and axial currents are the sum and the difference of the
light-like left- and the right- currents  of the Weyl spinors,
$j_a=j^{L}_a+j^{R}_a$ and $~{\cal J}_a=j^{L}_a-j^{R}_a~$.
There also exists a well-known set of algebraic identities between various
invariants of Dirac and Weyl spinors.
It is straightforward to check that
\begin{eqnarray}
{\cal J}_a j^{a}=0~,~~({\cal S}\pm i{\cal P})^2=
(\vec{L}\pm i\vec{K})^2~,\nonumber\\
j_a j^a=-{\cal J}_a{\cal J}^a =
2j^{L}_a~j^{Ra}={\cal S}^2+{\cal P}^2 >0~,
\label{eq:E2.5}\end{eqnarray}
where $K_i=M_{0i}$, and $2L_i=\epsilon_{ikm}M_{km}$.
These formulae relate the tensor densities, and they are
identically satisfied at every point.

\subsection{ The Dirac equation revisited.}

As was discussed in detail in paper \cite{paper1}, there is nothing
at our disposal that could have been used to create an equation for
the spinor field, except for the covariant derivative of a spinor
field. In a tetrad basis it was found to be $D \psi= D_a \psi d s^a =
(\partial_a \psi -\Gamma_a \psi) d s^a~,$ with the connection
$\Gamma_a$ given by equation (\ref{eq:E2.1}). The structure of this
spin connection is two-fold; its spin indices are being used to
parameterize rotations of the local tetrad basis by means of the
Pauli matrices, and its Lorentz index indicates the direction of
parallel transport. The first step is to parameterize the Lorentz
index $a$ by a spinor and thus convert this derivative entirely into
a spinor representation.  This conversion of two constituents of the
Dirac spinor is carried out by means of the matrix
$\alpha_a=(1,\rho_3\sigma_i)$; the left- and right- spinors are
Lorentz transformed {\em differently}.

Taking the linear relation, $u_\mu P^\mu=m$, as a classical prototype for
the equation of motion, we can write the following version of the Dirac
equation (and its conjugate),
\begin{eqnarray}
\alpha^a(\partial_a \psi -\Gamma_a \psi)+
im \rho_1 \psi=0~,\nonumber\\
(\partial_a \psi^+ - \psi^+ \Gamma_a^+ )\alpha^a-
im \psi^+ \rho_1=0~,
\label{eq:E2.6}\end{eqnarray}
with the spin connection (\ref{eq:E2.1}), which includes an
additional vector field $\aleph_a(x)$ that acts differently on
different spinor components. The differential operator of the Dirac
equation (\ref{eq:E2.6}) is Hermitian (symmetric), which is confirmed
by the conservation of vector (probability) current $j^\mu$, but this
does not necessarily means that it is a self-adjoint operator. In
fact, the axial potential can be so singular that, in general, it is
not self-adjoint unless a special set of requirements is imposed on
the axial field.

\subsection{ Conservation laws and equations of motion for the spin connections.}

The equations of motion (\ref{eq:E2.6}) allow one to
derive a number of identities. One of them,
\begin{eqnarray}
\nabla_\mu j^\mu=\nabla_\mu[\psi^+ \alpha^\mu \psi]
={1\over\sqrt{-{\rm g}}}\partial_\mu[\sqrt{-{\rm g}}
\psi^+ \alpha^\mu \psi]=0,~
\label{eq:E2.7}\end{eqnarray}
clearly indicates the conservation of the time-like probability current and
thus provides a definition of a scalar product in the space of Dirac spinor
fields as an integral over the three-dimensional space-like surface.
The second identity indicates that the axial current cannot be conserved,
\begin{eqnarray}
\nabla_\mu {\cal J}^\mu = \nabla_\mu[\psi^+\rho_3 \alpha^\mu \psi]=
2m\psi^+\rho_2 \psi=2m {\cal P}~,
\label{eq:E2.8}\end{eqnarray}
in full compliance with its space-like nature. Indeed, the
conserved current can only be time-like. The pseudoscalar
density is a measure of a dynamic interplay between the left-
and right- components of the Dirac spinor  (deviation
from a perfect parity-even configuration encoded in the conventional
Dirac equation \cite{Berest,Ryder}).

Next, we introduce a standard energy-momentum tensor, which is
conceived as a flux of the momentum,
\begin{eqnarray}
T^{\sigma}_{~\mu}=i ~\psi^+ \alpha^\sigma \overrightarrow{D_\mu} \psi~.
\label{eq:E2.9}\end{eqnarray}
Indeed, $\alpha^\sigma$ is the
quantum mechanical operator of the velocity and $D_\mu$ is a
prototype of the kinetic momentum $P_\mu=m u_\mu$. Its operator is
supposed to be Hermitian {\em and} self-adjoint. Hence, it may seem
reasonable to take $T^\sigma_{~\mu}$, in advance, in a manifestly
{\em real} form,
$2T^{\sigma}_{~\mu}=i \psi^+ [\alpha^\sigma \overrightarrow{D_\mu}
-\alpha^\sigma \overleftarrow{D^+_\mu}]\psi$. However, it is not safe
to have an operator that simultaneously acts in two adjoint spaces without
confidence that these two spaces coincide. One needs to
know beforehand  that the operator $D_\mu$ is self-adjoint,
which is not obvious because, in general, the axial potential in $D_\mu$
is singular. The condition (\ref{eq:E4.7}) for the
self-adjointness of $T^{\sigma}_{~\mu}$ will be derived
in Sec.~\ref{sec:Sec4}, and will render the {\em ad hoc} symmetrization
pointless.

It is straightforward to show (e.g., using the technique
of Ref.~\cite{Fock1}) that, by virtue of the
equations of motion, the following identity holds
\begin{eqnarray}
\nabla_\sigma T^{\sigma}_{~\mu}\equiv {1\over\sqrt{-{\rm g}}}
\partial_\sigma[\sqrt{-{\rm g}}T^{\sigma}_{~\mu}]-
\Gamma^\sigma_{\mu\nu}T^\nu_\sigma ~~~~~~~~~~~~\nonumber\\
=i~\psi^+ [D_\sigma D_\mu -D_\mu D_\sigma ]\psi-
2mg {\aleph}_\mu {\cal P}~.
\label{eq:E2.10}\end{eqnarray}
Here, the commutator of the covariant derivatives (the curvature tensor)
has the following tetrad representation,
$$[D_\sigma, D_\mu]=e_\sigma^a
[\partial_b\Gamma_a-\partial_a\Gamma_b +
\Gamma_a\Gamma_b-\Gamma_b\Gamma_a+C^c_{ab}\Gamma_c]e_\mu^b~,$$
where the structure constants $C^c_{ab}=\omega^c_{~ab}-\omega^c_{~ba}$
can be related to some Lie group associated with parallel transport.
An explicit computation shows that
\begin{eqnarray}
\nabla_\sigma T^{\sigma}_{~\mu}=
-eF_{\sigma\mu}j^\sigma
-g{\cal U}_{\sigma\mu}{\cal J}^\sigma \nonumber\\
+{i\over 4}\psi^+\alpha_\sigma
R^\sigma_{~\mu;cd}\rho_1\alpha^c\rho_1\alpha^d \psi
-2gm\aleph_\mu~{\cal P}~,
\label{eq:E2.10a}\end{eqnarray}
where the commutator of the covariant derivatives is expressed
in terms of two gradient invariant tensors (the field strengths),
$ F_{\mu\nu}=\partial_\mu A_\nu -\partial_\nu A_\mu$, and
${\cal U}_{\mu\nu}=\partial_\mu \aleph_\nu -\partial_\nu \aleph_\mu$,
and the Riemann curvature tensor,
$$R_{ab;cd}=\partial_b\omega_{cda}-\partial_a\omega_{cdb}
+\omega_{fca}\omega^f_{~db}-\omega_{fcb}\omega^f_{~da}
+C^f_{~ab}\omega_{fcd}~.$$
It is straightforward to show that the third term in Eq.~(\ref{eq:E2.10a})
can be transformed in the following way,
$$\psi^+\alpha^a
R^\sigma_{~\mu;cd}\rho_1\alpha^c\rho_1\alpha^d \psi =
 2R_{\mu\sigma}j^\sigma,$$
where $R_{ad}={\rm g}^{bc}R_{ab;cd}$ is the Ricci curvature tensor
\footnote{In his paper \cite{Fock1}, Fock emphasized the significance
of the presence of the Ricci tensor side by side with the field strength
tensor $F_{\mu\nu}$, deferring  any further discussion of this fact.}.

The next step is to convert Eq.~(\ref{eq:E2.10a}) into the divergence of one
common energy momentum tensor for all fields in the system. Hence,
we need the equations of motions for the fields $A_\mu$
and $\aleph_\mu$. The definition of the field tensors immediately yields
the first (without the  sources) couple of the Maxwell equations for the
field strengths $F_{\mu\nu}$ and ${\cal U}_{\mu\nu}$,
$$\nabla_\lambda \epsilon^{\sigma\lambda\mu\nu}F_{\mu\nu}=0~,~~~~
\nabla_\lambda \epsilon^{\sigma\lambda\mu\nu}{\cal U}_{\mu\nu}=0~. $$
The equations that interconnect fields and currents  were motivated by
the actual hydrogen spectra in the first paper.
The Lorentz invariant form of the Coulomb law for the field $A_\mu$ is
\begin{eqnarray}
\nabla_\sigma F^{\sigma\mu}=ej^\mu=e[\psi^+\alpha^\mu \psi]~,
\label{eq:E2.11}\end{eqnarray}
so that $F_{\mu\nu}$ is the massless gradient-invariant Maxwell
field which has the {\em probability current} as its source.
For the axial field $\aleph_\mu$ a plausible choice
is a massive  neutral vector field,
\begin{eqnarray}
\nabla_\sigma {\cal U}^{\sigma\mu}+M^2
\aleph^\mu=g {\cal J}^\mu,~~
M^2\nabla_\mu \aleph^\mu =
2gm{\cal P}~,
\label{eq:E2.12}\end{eqnarray}
where the second equation is the covariant derivative of the first one.

The Lorentz forces in Eq.~(\ref{eq:E2.10}), with the field tensors
$F_{\mu\nu}$ and ${\cal U}_{\mu\nu}$ in a familiar role,
prompt the same equations of motion, because these equations
allow one to present the  Lorentz force  as the divergence of
the energy-momentum tensor. We have
\begin{eqnarray}
ej^\sigma F_{\sigma\mu}=
\nabla_\lambda[F^{\lambda\nu}F_{\nu\mu}+
{1\over 4}\delta^\lambda_\mu F^{\rho\nu}F_{\rho\nu}]
=\nabla_\lambda \Theta^\lambda_{~~\mu}.~
\label{eq:E2.13}\end{eqnarray}
Using (\ref{eq:E2.12}) and (\ref{eq:E2.8}) one can transform
the Lorentz force of the axial field in (\ref{eq:E2.10}) into the
divergence of its energy-momentum tensor,
\begin{eqnarray}
g{\cal J}^\sigma {\cal U}_{\sigma\mu}
+g \aleph_\mu~(\nabla_\sigma {\cal J}^\sigma)=
\nabla_\lambda (\theta^\lambda_{~\mu}+t^\lambda_{~\mu})~,
\label{eq:E2.14}\end{eqnarray}
where
\begin{eqnarray}
\theta^\lambda_{~\mu}={\cal U}^{\lambda\nu}{\cal U}_{\nu\mu}+
{\delta^\lambda_\mu\over 4} {\cal U}^{\rho\nu}
{\cal U}_{\rho\nu}~,\nonumber\\
t^\lambda_{~\mu}=M^2\bigg(\aleph^\lambda \aleph_\mu -
{\delta^\lambda_\mu\over 2}\aleph^\rho \aleph_\rho\bigg)~.
\label{eq:E2.15}\end{eqnarray}
In both (\ref{eq:E2.11}) and in (\ref{eq:E2.12}) we followed the standard
convention, which defines the charge as the divergence of its electric
field, so  that the positive charge corresponds to
the positive flux of the electric field outside a surrounding surface.
With this convention, the energy components $\Theta^{00}$,
$\theta^{00}$ and $t^{00}$
of tensors (\ref{eq:E2.13}) and (\ref{eq:E2.15}) come out {\em
positive} exclusively because the coupling constants in the electron
spin connection (\ref{eq:E2.1}) were chosen {\em negative}.

For the reasons that will become clear later, the energy-momentum
tensor of the axial field is split into parts with and without
derivatives. The first part has the form of the Maxwell tensor. The
second one {\em will acquire} derivatives {\em after} the condition
of compactness (\ref{eq:E3.0}) is imposed on the Dirac field.
Putting Eqs.~(\ref{eq:E2.10})-(\ref{eq:E2.15}) together, one finds that
\begin{eqnarray}
\nabla_\lambda ( T^\lambda_{~~\mu} +\Theta^\lambda_{~~\mu}
+\theta^\lambda_{~~\mu}+t^\lambda_{~~\mu})=
{i\over 2}R_{\mu\sigma}j^\sigma=0.~
\label{eq:E2.16}\end{eqnarray}
The total energy-momentum of three interacting fields, $\psi$, $A_\mu$,
and $\aleph_\mu$ is conserved, which is an additional indication that
the system of equation of motion is self-consistent
(provided this system has a time-like Killing vector field and
the operator $D_a$ is self-adjoint.) The energy-momentum tensors
of both vector fields have {\em positive} energy densities, while
the energy of the Dirac field can have both signs. The second equation
in (\ref{eq:E2.16}), namely, $R_{\mu\sigma}=0$, will be
derived in Sec.~\ref{sec:Sec4}.

Formally, the equations of motion follow from the Lagrangian,
${\cal L}={\cal L}_D +{\cal L}_A +{\cal L}_\aleph~,$
where the Lagrangians of the three fields, $\psi$, $A_a$ and
$\aleph_a$ are,
$${\cal L}_D=i\psi^+ \alpha^\mu {D_\mu} \psi-m\psi^+\rho_1\psi$$
$${\cal L}_A=-{1\over 4}F^{\rho\nu}F_{\rho\nu},~~
{\cal L}_\aleph=-{1\over 4} {\cal U}^{\rho\nu}{\cal U}_{\rho\nu}-
{M^2\over 2}\aleph^\rho \aleph_\rho~,$$
and all fields at all space-time points are defined exclusively
with respect to a local tetrad basis.

\section{ Compact objects from Dirac spinors \label{sec:Sec3}}
\renewcommand{\theequation}{3.\arabic{equation}}
\setcounter{equation}{-1}

So far, we have dealt only with the definition of parallel transport,
equations of motion and the most primitive conservation laws. Almost
nothing has been said about the physical objects that are governed by
these equations.  Special relativity is built on two premises, the
light-like propagation of all fields that carry a signal and the
existence of inertial frames. The first one is readily implemented by
a spinor representation of the Lorentz group -- the two-component
Weyl spinors just map the light cone. The second one is more
difficult to implement because any inertial observer should have its
own proper time and  the solutions of the relativistic wave equations
are not easily localized to the extent so that they can serve as the
observers (rods and clocks) of special relativity. The electron with
a given momentum is just a plane wave! At the same time, all data
point to the fact that all truly localized interactions are due to
spinor fields. Therefore, the major problem is to constructively
identify the finite-sized spinor objects and a free space between
them. Without clarity at this point the entire concept of the
space-time continuum is vague and the idea of motion has no firm
physical footing.

The spinor field that has the axial vector field in its spin
connection seems to be more flexible than the original Dirac electron
and more suitable to approach this old problem. To give an idea of
how it might work, let us resort to the classical limit of a
relativistic point-like particle with the ``built in'' polarization
$\vec{\zeta}$ \cite{Weinberg}. If $\vec{\zeta}$ of the rest frame
is considered as a 4-vector, which is orthogonal to the 4-velocity,
$\zeta^\mu u_\mu=0$, then both  vectors are readily converted into
the vector fields {\em by means of~} their parallel transport,
$$ Du^\mu =(\partial_\nu u^\mu
+\Gamma^\mu_{\lambda\nu}u^\lambda)dx^\nu=0,$$
$$D\zeta^\mu =(\partial_\nu \zeta^\mu
+\Gamma^\mu_{\lambda\nu}\zeta^\lambda)dx^\nu=0~.$$
We can continue by taking $dx^\nu=u^\nu d\tau$ and end up with the
equations for the geodesic trajectory of a point-like particle and
the parallel transport of its polarization along this trajectory.
The vector $\vec{\zeta}$ is a classical prototype for the
{\em internal polarization} of the Dirac spinor field which is
represented by  various bilinear forms, like $\cal S$, $\cal P$,
etc. In the context of the Dirac equation, the vector $\vec{\zeta}$ is
associated with the components of the tensor $M_{\mu\nu}$ where the
left- and right- spinors are mixed by the matrices $\rho_1$ and
$\rho_2$. As long as we wish to treat the electron as a compact object
with a frozen-in polarization (including all its quantum numbers)
we have to {\em require} that these attributes are parallel transported,
\begin{eqnarray}
D{\cal S}=0~,~~~~D{\cal P}=0~,~~\ldots~,
\label{eq:E3.0}\end{eqnarray}
according to the definition  (\ref{eq:E2.3}).  The goal of this {\em
ansatz} is to put a compact object built entirely from spinor fields
(and their spin connections) into an inertial frame that accompanies
it. The spin connection of the Dirac field remains non-integrable,
$D\psi=(\partial_a-\Gamma_a)\psi ds^a\neq 0$. The ansatz
(\ref{eq:E3.0}) serves as an additional restriction on the possible
field $\aleph_a$ in close proximity of a stable spinor object.

\subsection{ Gordon's decomposition }

A list of attributes of a compact object that should be subjected to
the {\em ansatz} (\ref{eq:E3.0}) can be identified by means of the
so-called Gordon's decomposition. In many cases, it draws a clear
cut distinction between the convection and polarization currents. Using
spin connection (\ref{eq:E2.1}) and the algebraic relations,
$$\alpha^c\rho_1\alpha^d+\alpha^d\rho_1\alpha^c=2{\rm g}^{cd}\rho_1,$$
and
$$\alpha^c\rho_1\alpha^d-\alpha^d\rho_1\alpha^c=2\rho_1\Sigma^{cd},$$
where $\Sigma^{cd}=(1/2)[\gamma^c,\gamma^d]$, it is straightforward
to present the probability current, $j_a=\psi^+\alpha^a\psi$,
as follows,
\begin{widetext}
\begin{eqnarray}
j_a={i\hbar\over 2mc}\bigg\{\psi^+\rho_1
\stackrel{\leftrightarrow}{\partial_a}\psi -2ieA_a\psi^+\rho_1\psi
 + \partial_b(\psi^+ \rho_1\Sigma^{ab}\psi)
-2ig\aleph_b \psi^+\rho_1\Sigma^{ab}\rho_3\psi
+{1\over 4}
\omega_{cdb}\psi^+\rho_1[\Sigma^{ab},\Sigma^{cd}]\psi\bigg\}.~~~~~
\label{eq:E3.3}\end{eqnarray}
By virtue of the identity, $[\Sigma^{ab},\Sigma^{cd}]=2[{\rm g}^{ad}\Sigma^{bc}+
{\rm g}^{bc}\Sigma^{ad}-{\rm g}^{ac}\Sigma^{bd}-{\rm g}^{bd}\Sigma^{ac}],$
we have
$(1/4)\omega_{cdb} [\Sigma^{ab},\Sigma^{cd}]=
\Sigma^{ac}\omega_{bcb}-\Sigma^{bc}\omega_{acb}$,
and the probability current becomes
\begin{eqnarray}
j_a={i\hbar\over 2mc}\bigg\{\big[ \psi^+\rho_1
\partial_a\psi -(\partial_a\psi^+)\rho_1\psi
-2ieA_a\psi^+\rho_1\psi\big]
 + D_b(\psi^+ \rho_1\Sigma^{ab}\psi)\bigg\},~
\label{eq:E3.4}\end{eqnarray}
where
$ ~D_b[\psi^+ \rho_1\Sigma^{ab}\psi]=\nabla_b(\psi^+ \rho_1\Sigma^{ab}\psi)
-g\aleph_b \epsilon^{abcd}\psi^+\rho_1\Sigma_{cd}\psi~$
is the absolute derivative of the polarization tensor.
A similar representation is possible for the axial current,
${\cal J}_a=\bar{\psi}\gamma^a\gamma^5\psi$ ,
\begin{eqnarray}
{\cal J}_a={-\hbar\over 2mc}\bigg\{
\big[\psi^+ \rho_2 \Sigma^{ab} (\overrightarrow{\partial_b}-\Omega_b)\psi
-\psi^+ (\overleftarrow{\partial_b}-\Omega_b)\rho_2 \Sigma^{ab}\psi
 +2ieA_b(\psi^+ \rho_2 \Sigma^{ab}\psi)\big]
+  D_a(\psi^+\rho_2\psi)\bigg\}.
\label{eq:E3.4a}\end{eqnarray}
\end{widetext}

The original ``diagonal'' representation of the probability current
$j_a$ (that does not mix left- and right- spinors) is traded for a
sum (\ref{eq:E3.4}) of the ``off-diagonal'' terms (where these
spinors are mixed). The first three terms (in brackets) is the
convection current of the Schr\"{o}dinger equation for a
structureless particle, the last term is due to the internal
polarization that is transported with the particle. In the
axial current, the convection ``drags'' the density of internal
polarization, while the pseudoscalar density ${\cal P}$ is parallel
transported with the particle, and it cannot change unless this
particle is subjected to an external field or decays. Quite
remarkably, the convective derivative includes only the Maxwell
field, which acts on a Dirac particle as a whole, while the parallel
transported polarization terms are affected only by the axial field,
which differentiates between the right and left spinor components.

\subsection{ Axial field of a compact Dirac particle}

The ansatz (\ref{eq:E3.0}) serves as an additional condition on the spin
connection $\rho_3\aleph_a(x)$ of the Dirac field, which
makes it compatible with the existence of a freely moving stable
spinor particle. With an {\em a priori} form (\ref{eq:E2.1}) of the spin
connection, it can be cast into the following coordinate form,
\begin{eqnarray}
D_\mu {\cal S}=\nabla_\mu {\cal S}+2g\aleph_\mu {\cal P}=0,\nonumber\\
D_\mu {\cal P}=\nabla_\mu {\cal P}-2g\aleph_\mu {\cal S}=0.
\label{eq:E3.6}\end{eqnarray}
(If there were no axial field $\aleph_\mu$, then we would have only
one condition, $\nabla_\mu M^{\lambda\nu}=0$ (cf. Eq.~(\ref{eq:E3.3})).
As one can anticipate, the electric polarization current in free
space is zero, $~j^\nu_{polariz}=\nabla_\lambda M^{\lambda\nu}=0$.)

Acting on Eqs.~(\ref{eq:E3.6})  by $\nabla_\mu$, and
excluding the first derivatives we obtain the equations,
\begin{eqnarray}
\Box{\cal S}+ 4g^2\aleph^2{\cal S}
= -(4g^2m/ M^2)~{\cal P}^2, \nonumber\\
\Box {\cal P}+
4g^2\aleph^2{\cal P}= (4g^2m/M^2)~{\cal S}{\cal P}.~~
\label{eq:E3.7}\end{eqnarray}
Eqs.~(\ref{eq:E3.7}) do not  depend on the Maxwell field $A_\mu$ and
are the relativistic wave equations. This means that the propagation
of static polarization of an isolated compact object can be reduced
to a Lorentz transformation of the proper fields, which are
completely defined in its rest frame. The densities ${\cal S}$,
${\cal P}$,... become the {\em effective fields} solely because
we wish a compact spinor object to have a co-moving frame!
Then {\em for any short-scale dynamics that might take place in this
frame, the translational invariance must be explicitly broken}. From
this viewpoint, the function $~\aleph^2=\aleph_0^2-\vec{\aleph}^2~$
in Eqs.~(\ref{eq:E3.7}) stands for a scalar potential and it
determines if the ``center of a particle'' is attractive or repulsive
for the fields of polarization and, eventually, for the other
particles. The densities ${\cal S}$ and ${\cal P}$, are the simplest
possible ones. Any of them can serve as a mass term and provide the primary
binding of two Weyl spinors.

As one can anticipate, the ansatz (\ref{eq:E3.0}) indeed affects
the possible form of the connection $\aleph^a$ (and only of $\aleph^a$).
Instead of being an arbitrary external vector field  it becomes a
functional $\aleph_a\{{\cal S},{\cal P}\}$ of spinor forms:
from Eqs.~(\ref{eq:E3.6}) it immediately
follows that
\begin{eqnarray}
 2g \aleph_\mu\{{\cal S},{\cal P}\}  = \partial_\mu [\arctan({\cal P}/{\cal S})]
 = -\partial_\mu\Upsilon\{{\cal S},{\cal P}\}.
\label{eq:E3.9a}\end{eqnarray}
Alternatively, one can use Eqs.~(\ref{eq:E3.6}) to show that
${\cal S}\partial_\mu{\cal S}+{\cal P}\partial_\mu{\cal P}= 0$.
Hence, the scalar function~ ${\cal R}^2={\cal S}^2 +{\cal P}^2$,
is the first integral of these two equations (for a toy model
of the localized solution with this property see Sec.~\ref{sec:Sec5} ).
This had to be expected because the current $j^a$ is time-like and its
conservation means that $D_aj^a=0$. Since ${\cal S}$ and
${\cal P}$ are real functions, we may look for a solution of the form
$${\cal S}={\cal R}\cos\Upsilon~, ~~~~~{\cal P}=-{\cal R}\sin\Upsilon~,$$
which yields the same equation as Eq.~(\ref{eq:E3.9a}),
\begin{eqnarray}
 2g \aleph_\mu  = - \partial_\mu \Upsilon~.
\label{eq:E3.9}\end{eqnarray}
Since ${\cal R}^2=j^aj_a$ is the squared probability current, the
positive ${\cal R}$ is a natural measure for {\em
localized} spinor matter.  {\em The potential function $\Upsilon(x)$
remains a distinct characteristics of the Dirac field even in that
part of space where ${\cal R}\to 0$.} For perfectly stable spinor
matter, the axial field $\aleph_\mu$ becomes very simple; it is
completely defined by a single scalar function\footnote{
This very much resembles general relativity: the affine connection
$\Gamma^\lambda_{\mu\nu}$ has, in general, $4^3=64$ components.
However, if it defines a parallel transport of four vector fields,
e.g. a tetrad, then its 64 components can be expressed via only
16 functions.}. Its curvature tensor vanishes,
${\cal U}^{\sigma\mu}=0$. Furthermore, if the conditions
(\ref{eq:E3.0}) are exact, then the Eqs.~(\ref{eq:E2.15}) become
 \footnote{If
(\ref{eq:E3.0}) is not perfectly satisfied, then one may {\em think
of} a not perfectly stable cluster where the field $\aleph_\mu$ is
not an exact gradient of $\Upsilon(x)$, and where some remnant of the
field strength ${\cal U}^{\sigma\mu}$ may interfere with
$F^{\sigma\mu}$. Most likely, this will lead to the loss of
the normalizeable states, etc. However, this seems
to be a natural way to discover an entire spectrum of {\em unstable}
elementary particles as the excitations of the Dirac field. }
\begin{eqnarray}
\theta^\lambda_{~\mu}=0,~~~~
t_{\lambda\mu}={M^2\over 4g^2}
\bigg(\partial_\lambda\Upsilon \partial_\mu\Upsilon -
{g_{\lambda\mu}\over 2}\partial_\rho\Upsilon
\partial^\rho\Upsilon \bigg).~~
\label{eq:E3.10}\end{eqnarray}
This is the energy-momentum  tensor of the massless scalar field
$\Upsilon(x)$ with a positively defined Hamiltonian. Thus, the full
field $\aleph$ has lost its transverse part but it still remains a
dynamical object. The equation for this field can be found by
computing the covariant divergence of Eq.~(\ref{eq:E3.9}) and
expressing the divergence of the axial field via the pseudoscalar
density,
$M^2\nabla_\mu \aleph^\mu=2gm{\cal P}=-2gm {\cal R}\sin\Upsilon$.
We find that
\begin{eqnarray}
\Box~ \Upsilon(x)-{4g^2m\over M^2}~{\cal R}(x)\sin\Upsilon(x)=0~.
\label{eq:E3.12}\end{eqnarray}
Within a domain where ${\cal R}(x)$ is a constant, this is a
well-known sine-Gordon equation which has the soliton solutions.
This equation is classical, {\em the Planck constant is gone};
therefore, the field $\Upsilon(x)$ does not vanish in the classical limit.
In the case of a variable source, the variations of $\Upsilon(x)$
propagate according to the d'Alembert equation.
It is not clear if this equation has static solutions at all.
However, if such solutions do exist, then
of particular interest is the case of spherical symmetry.
If the exterior of the domain $r<r_{max}$ is empty space where
${\cal R}^2=0$, then the function $\Upsilon$ is a solution to
the external problem of the Laplace equation\footnote{
Since ${\cal R}^2= j^aj_a$, we have ${\cal R}=0$ (and must
call this space empty) when the density of probability current in
it is a light-like vector. In this case, the arguments that
have led to the ansatz (\ref{eq:E3.0}) are not valid. The two-component
spinor field cannot form a compact object. It cannot
 be a source of the field $\Upsilon$. }.
 Therefore, at $r>r_{max}$ we have
\begin{eqnarray}
\Upsilon(r)= - {1\over r}~{4g^2 m\over M^2}~
\int_0^{r_{max}}{\cal P}(r)r^2~dr~.
\label{eq:E3.14}\end{eqnarray}
Now, the most convincing argument in favor of a strong localization comes
from the Dirac equation itself, where the potential
$-\partial_r\Upsilon \sim -1/r^2$ enters the radial component
$ig\rho_3\aleph_r$ of the spin connection  and $\aleph_a(x)$ is the
remaining (less singular)
part of the axial field which is not regulated by the ansatz and has
a non-vanishing field tensor ${\cal U}_{ab}$,
\begin{eqnarray}
\alpha^a \bigg\{\partial_a
+ieA_a(x)+ig\rho_3\aleph_a(x)
+{i\over 2}~\rho_3{\partial\Upsilon(x)\over \partial x^a}\nonumber\\
 -\Omega_a(x)\bigg\}\psi +im \rho_1\psi=0.~~
\label{eq:E3.15}\end{eqnarray}

The most singular potential  $~\partial_a\Upsilon~$ brings in the
term $\propto 1/r^4$ into an equivalent Schr\"{o}dinger equation
signaling an advent of the ``falling onto the centre'' phenomenon
(and there is no threshold condition like $Z>137$ in the Coulomb
field). An obvious conjecture is that this super-critical binding is
likely to initiate a (rich) spectrum of (quasi)bound states where the
Dirac field has a {\em negative energy}.

We may generalize this observation by realizing that, by a formal
integration of Eq.~(\ref{eq:E3.12}) and expressing $\Upsilon(x)$ in
the Dirac equation (\ref{eq:E3.15}) as an integral of ${\cal P}$, we
convert the latter into a non-linear integral-differential equation
for the spinor field $\psi$. In this equation, the mass parameter $m$
will no longer be an arbitrary number.

One of the solutions of the coupled equations (\ref{eq:E3.12}) and
(\ref{eq:E3.15}) is obvious. This is the solution with $m=0$
(formally) or, equivalently, with ${\cal R}(x)=0$ (physically). The
probability current for this spinor mode is light-like. Such a mode
can have only one, left- or right- spinor component, and it
completely detaches from the field $\Upsilon(x)$ (except for the
obvious effect of the metric that will be derived in the next
section).

\section{ Axial forces and gravity. \label{sec:Sec4}}
\renewcommand{\theequation}{4.\arabic{equation}}
\setcounter{equation}{-1}

The requirement (\ref{eq:E3.0})
that the internal polarization densities like ${\cal S}$ and ${\cal P}$
must be frozen into a stable Dirac particle has limited the form of the
axial field in the spin connection to a gradient of the potential function
$\Upsilon$. This step has also led to a non-linear system of
equations (\ref{eq:E3.12})-(\ref{eq:E3.15}) that can
yield a hierarchy of scales. Thus, this system can have localized
solutions which represent  compact clusters of the spinor field.

\subsection{Localization of energy and the force of inertia}

Any compact object must also localize its energy-momentum along
the world line of a particle. Therefore we are forced to augment the
ansatz (\ref{eq:E3.0}) by a new element,
\begin{eqnarray}
D_\sigma T^\sigma_\mu=0~.
\label{eq:E4.0}\end{eqnarray}
This is an expression of the fact that the kinetic 4-momentum
is parallel-transported with the particle and does not change when
it is moving in free space. By the definition, we have
\begin{eqnarray}
D_\sigma T^\sigma_{~\mu}\equiv i
 (\psi^+ \overleftarrow{D^+_\sigma}\alpha^\sigma
 \overrightarrow{D_\mu} \psi +
\psi^+ \alpha^\sigma \overrightarrow{D_\mu}
\overrightarrow{D_\sigma} \psi) =0~,
\label{eq:E4.1}\end{eqnarray}
which yields the equation,
\begin{eqnarray}
{1\over\sqrt{-{\rm g}}}\partial_\sigma(\sqrt{-{\rm g}}T^\sigma_{~\mu})=
i \psi^+ \alpha^\sigma[D_\mu D_\sigma -D_\sigma D_\mu]\psi~.
\label{eq:E4.2}\end{eqnarray}
It has to be compared with the identity (\ref{eq:E2.10}) that follows from the
equations of motion,
\begin{eqnarray}
 {1\over\sqrt{-{\rm g}}} \partial_\sigma[\sqrt{-{\rm g}}T^{\sigma}_{~\mu}]
=\Gamma^\sigma_{\mu\nu}T^\nu_{~\sigma}~~~~~~~~~~\nonumber\\
+i\psi^+ [D_\sigma D_\mu -D_\mu D_\sigma ]\psi
-2mg{\aleph}_\mu {\cal P}.
\label{eq:E4.3}\end{eqnarray}
These two equations coincide if the force of inertia and the external
force from axial field $\aleph$ are equal, i.e.,
\begin{eqnarray}
 \Gamma^\sigma_{\mu\nu}T^\nu_{~\sigma} =
2m g {\aleph}_\mu {\cal P}
=- m  {\cal P}~\partial_\mu \Upsilon ~.
\label{eq:E4.4}\end{eqnarray}
While the first part (\ref{eq:E3.0}) of the ansatz  simplified
the spin connection ${\aleph}_\mu$ to the gradient of a scalar
function, the second part
(\ref{eq:E4.0}) specifies the affine connection. Indeed,
the Ricci rotation coefficients $\omega_{cda}$ that enter the
spin connection $~\Omega_a$ are related to the Christoffel
symbols $ \Gamma^\sigma_{\mu\nu}$  through the covariance
of the tetrad vectors, $D_be_\mu^a=0$.
The simplest form of Eq.~(\ref{eq:E4.4}) is
$T^{00}\partial_i {\rm g}_{00}=2mc^2{\cal P}\partial_i\Upsilon $,
which immediately leads to
$${\rm g}_{00}=1+2{m{\cal P}\over T_{00}}\Upsilon \to
1+{2\Upsilon_{Newton}\over c^2}~.$$
Thus, we have {\em derived} the key formula, which is traditionally
considered as a  {\em phenomenological} basic of general
relativity, and thus the Newtonian form (\ref{eq:E3.14}) of the field
$\aleph$ at large distances is not a mere coincidence. The
field $\aleph_\mu$, by all its properties, is indistinguishable
from the gravitational field, which is also locally equivalent to
the field of the inertia forces. Indeed, in a local co-moving
(geodesic) reference frame, where the Christoffel connections
$\Gamma^\sigma_{\mu\nu}$ vanish, the potential $\Upsilon$ must
become constant (up to small second order corrections).
Furthermore, the pseudoscalar density ${\cal P}$ of a free Dirac
plane wave is zero. The spinor field of a plane-wave electron
is perfectly parity-even (see \cite{Berest,Ryder}) , and its
``gravitational mass''
${\cal P}$ is effectively switched off (it cannot be switched off
globally as long as {\em two} bodies interact and accelerate).

Eq.~(\ref{eq:E4.4}) explicitly states that for a small body, which
moves under the action of the axial field, it is possible to choose
such a parameterization of the space-time coordinates (the metric
tensor $g_{\mu\nu}(x)$) that this body will move along the geodesic
lines of {\em this} metric. In such a parameterization, the affine
connections $\Gamma^\sigma_{\mu\nu}$ take a role of the forces of
inertia that locally compensate the physical axial forces. This
substitution is impossible for the most general axial and spinor
fields. The latter ones must form stable compact objects, which
renders the axial-Newton's gravitational field $\Upsilon$ classical.
Surprisingly enough, this predetermines a possible form of the metric
tensor. The ansatz (\ref{eq:E3.0}),  (\ref{eq:E4.0}) completely
breaks up for unstable objects.

\subsection{Self-adjointness and the Einstein field equations}

 In the previous sections, keeping focus on the physical
picture, we manipulated the operators of the Dirac equation and
the energy-momentum tensor without confidence that these operators
even exist. At the same time, the axial field $\aleph$ was shown to
be singular, which means that the spaces of spinor functions $\psi$
and $\psi^+$ can be different. Therefore, we have to demand that the
above operators are self-adjoint.

The differential operator $i D_\mu$ of the Dirac  equation is
Hermitian, which leads to the conservation of the probability
current $j^\mu$. It is self-adjoint if
\begin{eqnarray}
i \int \partial_\sigma
(\sqrt{-{\rm g}}\psi^+ [\alpha^\sigma \overrightarrow{D_\mu} +
 \overleftarrow{D^+_\mu} \alpha^\sigma ] \psi)
 d^3 {\vec x} dx^0 \nonumber\\
=i\int \partial_\sigma \nabla_\mu
(\sqrt{-{\rm g}}\psi^+\alpha^\sigma\psi) d^3 {\vec x} dx^0 =0.
\label{eq:E2.10b}\end{eqnarray}
This Lorentz invariant equation encodes two requirements.
First, that the self-adjointness is defined with the scalar product
as an integral over a space-like surface. Second, that this condition
is the same for all space-like surfaces.

The Dirac equation (\ref{eq:E3.15}) can have stable compact
solutions with real energies only when the operator of
energy-momentum is self-adjoint. Then the solutions
of the Dirac equation and its adjoint belong
to the same space (e.g., have the same spectra of energies). This
condition has to be parallel-transported with the compact object that
``owns'' this spectrum, and the covariant form of this requirement is
\begin{eqnarray}
D_\sigma \big( \psi^+ [ \alpha^\sigma \overrightarrow{D_\mu} +
 \overleftarrow{D^+_\mu} \alpha^\sigma ] \psi\big)=
 \psi^+ [\alpha^\sigma \overrightarrow{D_\mu}  +
 \overleftarrow{D^+_\mu} \alpha^\sigma]
 \overrightarrow{D_\sigma}\psi\nonumber\\
 + \psi^+ \overleftarrow{D^+_\sigma}[\alpha^\sigma \overrightarrow{D_\mu}  +
 \overleftarrow{D^+_\mu} \alpha^\sigma ] \psi=0,~~
\label{eq:E4.5}\end{eqnarray}
which can be identically transformed into
\begin{eqnarray}
 \partial_\sigma( \sqrt{-{\rm g}}
[ \psi^+ \alpha^\sigma \overrightarrow{D_\mu} \psi +
\psi^+ \overleftarrow{D^+_\mu} \alpha^\sigma  \psi])\nonumber\\
+R_{\mu\sigma} ~ \sqrt{-{\rm g}}\psi^+ \alpha^\sigma\psi=0,~
\label{eq:E4.6}\end{eqnarray}
then integrated over the space-time domain, and compared with the
condition (\ref{eq:E2.10b}). This leads to a conclusion that
the Dirac field can have a self-adjoint Hamiltonian only if it
lives in a {\em free space} (as it is understood in general
relativity). Since $j^\mu\neq 0$, we must have
\begin{eqnarray}
R_{\lambda\sigma}=0~.
\label{eq:E4.7}\end{eqnarray}
In general relativity theory, this is Einstein's field equation.
 The Dirac spinor field, taken as matter,
not only leads to this equation, but allows one to derive the principle
of inertia and establish the identity between the axial and gravitational
fields.

Several remarks are in order. \\
(i) The ansatz (\ref{eq:E3.0}) and (\ref{eq:E4.0}) together with equation
(\ref{eq:E4.4}) are the spinor equivalent of the Fermi-Walker transport
of tensor fields \cite{FermiWalker} which are confined to the closest
vicinity of the particle's world line.\\
(ii) The origin of Eq.~(\ref{eq:E4.7}) in the context of this work
clearly supports Einstein's ultimate opinion that the
energy-momentum tensor does not represent the gravitating matter
adequately. It also supports Einstein's later attempts to
identify matter with singularities of the field equations
(\ref{eq:E4.7}) \cite{EI}. The major result was that, even
considered separately, the field equations for the metric of free
space have {\em singularities that are moving according to the laws
of motion of classical particles}. The principle of motion along a
geodesic line  follows from the field equations and is not an
independent principle.\\
(iii) Equations (\ref{eq:E3.12}),
(\ref{eq:E3.15}) and (\ref{eq:E4.7}), (\ref{eq:E4.4}) clearly
support the image of particles as moving singularities of the {\em
interacting Dirac and gravitational fields}. The gravitational force
appears to be a spin connection of the Dirac field. In fact, this
system guarantees that the theory will be protected from {\em
mathematical singularities} which inevitably show up when the
nonlinear Einstein's equations (\ref{eq:E4.7}) are solved
independently \cite{EI}. A striking agreement between the character
of motion of singular domains of the Einstein and Dirac fields seems
to be an indication that general relativity naturally requires
matter in the form of the Dirac field. No other fields can
provide the degree of localization, which is necessary for such a
coincidence. The complementarity of these two approaches indeed solves the
problem of motion as it was first posed by Einstein.

\section{Localization of the Dirac field and the gravitational
mass. \label{sec:Sec5}}
\renewcommand{\theequation}{5.\arabic{equation}}
\setcounter{equation}{0}

In this section, the toy model of a compact Dirac particle will be
worked out. This model captures some essential properties of the yet
unknown exact solutions of the non-linear system of
Eqs.~(\ref{eq:E3.12}),(\ref{eq:E3.15}) and (\ref{eq:E4.7}). The model
employs a symmetric closed configuration of the Dirac field, which
corresponds to the previously conjectured \cite{paper1} possibility
that the Dirac field is radially polarized in the internal geometry
of a sphere. It is an eigenstate of the operator $\sigma_3$
associated with the radial tetrad vector $e^3_a$, so that the Dirac
spinor is parallel-transported along a sphere.
 One of the possible prototypes of such a state is the fully occupied
electron shell of a noble gas, which is so symmetric that
none of the electrons can be assigned individual quantum numbers
associated with the angular momentum.  Another example is the linear
oscillatory orbit of the Bohr-Sommerfeld model. The only topologically
distinctive direction is the radial one, inward or outward.

Strictly speaking, considering this static model more seriously,
one has to use the Schwarzchild solution as the uniquely defined
(according to the Birkhoff theorem) metric background. According to
Eq.~(\ref{eq:E4.4}), the gravitational radius of this metric
is defined by the distribution of the pseudoscalar density
${\cal P}$, the peak localization of the Dirac field, rather than by
its energy (inertial mass). Any attempt to incorporate these
elements leads to the issue of the very existence of the
meaningful (measurable) metric relations near the Planck scale and,
possibly, even of the origin of the quantum-mechanical ensembles
\cite{tHooft}, which is beyond the scope of this paper.

Any extension of this model will also include modes with angular
dependence and, perhaps, some relevant quantum numbers. These modes
can have special names, e.g. $\{R,G,B\}$, but these names cannot be
given to plane waves -- they are meaningful only in the internal
space of closed configurations. As it was pointed out in paper
\cite{paper1}, in the presence of the axial field the angular
momentum is not conserved even in a perfectly spherical geometry.
This fact points to the possibility that in the course of the
formation of a closed spinor configuration some of the polarization
degrees of freedom undergo a transition from the ``normal'' world (of
the directions to distant stars) into a compactified internal space.
This internal space (invisible from the outside) is very likely to
have a non-Abelian symmetry group and, possibly, a local gauge
structure associated with it. Therefore, the well known non-Abelian
gauge theories may have a natural realization within this scheme. The
radial modes of these configurations can also have special names,
e.g., $\{u,d,\ldots\}$. The study of the emerging non-Abelian
structure, similar to theat which was pioneered in
Ref.~\cite{Wahlquist}, is currently underway.

From Eqs.~(4.11) and (4.14) of the paper \cite{paper1}, we have
two systems of equations for two topologically distinct modes

Equations for the modes with outward polarization are
\begin{eqnarray}
~[E_{\uparrow}-{Ze^2\over r} - g\aleph_r - i\partial_r] f_{\uparrow}=
m h_{\uparrow},\nonumber\\ ~[E_{\uparrow}-{Ze^2\over r} - g\aleph_r +
i\partial_r]h_{\uparrow}= m f_{\uparrow}.
\label{eq:E5.1}\end{eqnarray} where the functions $f$ and $h$ include
the factor $~(-{\rm g})^{1/4}=r\sqrt{\sin\theta}~$ from the Jacobian,
so that the measure of the volume integration is $~dr d\theta d\phi$
(at this instance only we include in these equations the attractive
Coulomb potential of the point-like charge $~+Ze~$, as a reference).
For inward modes we have the same system with the opposite sign of
$\aleph_r$. Introducing the new functions,
$$\sqrt{2}F=f+h~~~~{\rm and}~~~~i\sqrt{2}G=f-h,$$
we obtain equations with real coefficients,
\begin{eqnarray}
F'_{\uparrow}= [m+ E_{\uparrow} -g\aleph_r ] G_{\uparrow},~~
G'_{\uparrow}= [m- E_{\uparrow} +g\aleph_r ] F_{\uparrow},~~~
\label{eq:E5.3}\end{eqnarray}
where $F'=dF/dr$.  For both modes (and assuming real
solutions) we have the identity
\begin{eqnarray}
(F^2+G^2)'=4mFG.
\label{eq:E5.5}\end{eqnarray}
(In our simple case, it can be traced back to
$~\nabla_\mu {\cal J}^\mu =2m{\cal P}~$ of
Eq.~(\ref{eq:E2.8}) and  $~{\cal J}^2=-j^2~$ of Eq.~(\ref{eq:E2.5}) ).
This identity allows one to find the volume integral of pseudoscalar density by
means of the relations,
\begin{eqnarray}
{\cal P}={\cal P}_{\uparrow}+{\cal P}_{\downarrow},~~~~~~
\sqrt{-{\rm g}}~{\cal P}_{\uparrow\downarrow}=
\mp 2F_{\uparrow\downarrow}G_{\uparrow\downarrow}~.
\label{eq:E5.6}\end{eqnarray}
Let us think of a localized distribution of the Dirac field as a source of
a static potential $\aleph_r^{(0)}(r)$.
Then, according to Eq.~(\ref{eq:E3.14}),
\begin{eqnarray}
g\aleph_r^{(0)\uparrow}(r)=-{1\over r^2}~{2m g^2\over M^2}~
\int_0^{r_{max}}{\cal P}_{\uparrow}(r)r^2~dr\nonumber\\
={1\over r^2}~{ g^2\over M^2}\int_0^{r_{max}}
(F_{\uparrow}^2+G_{\uparrow}^2)'dr . \nonumber
\end{eqnarray}
Since the object is localized, there is no contribution from the upper
limit
\begin{eqnarray}
g\aleph_r^{(0)\uparrow}(r)
=-{1\over r^2}{ g^2\over M^2} R_{\uparrow}(0)~
=-{{\cal Q}_{\uparrow}\over r^2},~
\label{eq:E5.7}\end{eqnarray}
where $R(r)drd\Omega=r^2{\cal R}(r)drd\Omega$ is the physical
probability density. (For a
$\downarrow$-center we obviously have to change sign in the r.h.s.)
If this probability is normalized to unity within a sphere of radius
$r_{max}$, then $R(0)\sim1/r_{max}$ and, according to
Eq.~(\ref{eq:E5.17}), this estimate can be very accurate.
This provides a simple formula,
\begin{eqnarray}
2m\int{\cal P}_{\uparrow}dV \approx R(0)~.
\label{eq:E5.7a}\end{eqnarray}
 Physically, this quantity represents the gravitational mass in
Eqs.~(\ref{eq:E3.14}),  (\ref{eq:E4.3}), and (\ref{eq:E4.4})
clearly indicating that, at the microscopic level, the strength of
the axial-Newton interaction is proportional to the {\em peak of
localization} of the Dirac field at the gravitating centers.

The next step is to solve Eqs.~(\ref{eq:E5.1})
with these external sources.  For the sake of definiteness, let us take
(\ref{eq:E5.7}) as an external field. Then
Eqs.~(\ref{eq:E5.1}) and (\ref{eq:E5.3})  with $\uparrow$-center are
\begin{eqnarray}
[E_{\uparrow} +{{\cal Q}_{\uparrow}\over r^2} - i\partial_r] f_{\uparrow}\!=
m h_{\uparrow},~
[E_{\uparrow} +{{\cal Q}_{\uparrow}\over r^2} + i\partial_r]h_{\uparrow}\!=
m f_{\uparrow},\nonumber\\
F'_{\uparrow}= [m+ E_{\uparrow}+
{{\cal Q}_{\uparrow}\over r^2} ] G_{\uparrow},~
G'_{\uparrow}= [m- E_{\uparrow} -
{{\cal Q}_{\uparrow}\over r^2} ] F_{\uparrow},~~~~
\label{eq:E5.9}\end{eqnarray}
Their primary form (the upper line) indicates that we have the
Dirac equation with a {\em repulsive} singular potential.
The true degree of singularity is seen from the second order
equation, i.e.,
\begin{eqnarray}
f'' +\bigg[{2i{\cal Q}\over r^3}+
\bigg(E-{{\cal Q}\over r^2}\bigg)^2-m^2 \bigg]f =0~,
\label{eq:E5.9a}\end{eqnarray}
(which is similar to the one studied in Ref.~\cite{Regge}).

Eqs.~(\ref{eq:E5.9})  depend on three dimensional parameters, which
are the mass parameter $m$, the energy $E$ and the intensity of the
central charge ${\cal Q}$ (with the dimension $m^{-1}$). Looking for
the simplest possible object, I shall retain only two of them (by
taking $E=m$ or $E=-m$). This will render Eqs. (\ref{eq:E5.9}) easily
solvable. Indeed, for the most interesting case, $E=-m<0$, we have
\begin{eqnarray}
G'_{\uparrow}= \bigg[2m-{{\cal Q}_{\uparrow}\over r^2}\bigg]F_{\uparrow},
~~r^2 F'_{\uparrow}= {\cal Q}_{\uparrow}G_{\uparrow}.~~
\label{eq:E5.10}\end{eqnarray}
Differentiating the second equation, using the first one, and
changing the independent variable to $y=1/r$, we obtain
\begin{eqnarray}
{d^2 F_{\uparrow}\over dy^2}+\bigg({\cal Q}_{\uparrow}^2-
{2m{\cal Q}_{\uparrow}\over y^2}\bigg)F_{\uparrow}=0~,~~
G_{\uparrow}={-1\over {\cal Q}_{\uparrow}}{dF_{\uparrow}\over dy}.~~
\label{eq:E5.14}\end{eqnarray}
The solution that vanishes when $y\to 0$ ($r\to\infty$) at
all negative energies is
\begin{eqnarray}
F_{\uparrow}(y)= C_\nu y^{1/2}J_\nu({\cal Q}_{\uparrow}y)
= {C_\nu\over r^{1/2}}~J_\nu\big({{\cal Q}_{\uparrow}\over r}\big)~,
\nonumber\\
\nu^2={1\over 4}+2m{\cal Q}_{\uparrow}>0~.
\label{eq:E5.15}\end{eqnarray}
The normalization integral, $\int[F^2+G^2]dr$,
converges (at $r\to\infty$) only when $~\nu>1~$, and the
normalization coefficient is
$$C_\nu=\bigg({16\nu(\nu^2-1)\over 3(4\nu^2 -1)}\bigg)^{1/2}~.$$
The behavior of this solution as $r\to 0$ is noteworthy:
\begin{eqnarray}
F\approx C_\nu\sqrt{2\over\pi{\cal Q}_{\uparrow}}
\cos{{\cal Q}_{\uparrow}\over r}~,~~
G\approx C_\nu\sqrt{2\over\pi{\cal Q}_{\uparrow}}
\sin{{\cal Q}_{\uparrow}\over r},~~
\label{eq:E5.17}\end{eqnarray}
so that the probability density $~F^2+G^2~$ remains surprisingly
constant within the range of validity of the asymptotic formula for
the Bessel functions. These normalizeable bound states occupy the
range of energies $-\infty<E_{\uparrow}<-E_*=-3/(8{\cal Q}_{\uparrow})$,
and their energy spectrum is {\em continuous}. In the energy interval
$E_*<E<0$, the solution still goes to zero as $r\to\infty$,
but not sufficiently fast as to
be a true bound state without an additional cutoff (which is easily
provided, e.g., by an attractive Coulomb potential).

In exactly the same way, we consider the case $E_{\uparrow}=m>0$,
which leads to the equations
\begin{eqnarray}
{d^2G_{\uparrow}\over dy^2}+\bigg({\cal Q}_{\uparrow}^2+
{2m{\cal Q}_{\uparrow}\over y^2}\bigg)G_{\uparrow}=0~,~
F_{\uparrow}={1\over {\cal Q}_{\uparrow}}{dG_{\uparrow}\over dy}.~~
\label{eq:E5.11}\end{eqnarray}
The only solution to this equation which vanishes when $y\to 0$
($r\to\infty$)  is
\begin{eqnarray}
G_{\uparrow}(y)= C_\mu y^{1/2}J_\mu({\cal Q}_{\uparrow}y),~~~
\mu^2={1\over 4}-2m{\cal Q}_{\uparrow}>0~.~~~
\label{eq:E5.12}\end{eqnarray}
At $r\to 0$ ($y\to\infty$) this solution also infinitely oscillates, which
indicates a ``fall onto a centre''.
Therefore, this  is a solution confined around a seemingly
repulsive core and is a quasi-bound state when
$0<E_{\uparrow}<E_\#=1/(8{\cal Q}_{\uparrow})$. Since
$\mu^2<1$, it is not normalizeable as a true bound state.
In the opposite case,
when $E_{\uparrow}>E_\#$, we have $\lambda^2=-\mu^2>0$. A general
solution,
\begin{eqnarray}
G_{\uparrow}(y)= C y^{1/2}[e^{i\beta}J_{i\lambda}({\cal Q}_{\uparrow}y)
+e^{-i\beta}J_{-i\lambda}({\cal Q}_{\uparrow}y)]~,
\label{eq:E5.13}\end{eqnarray}
is not localized, unless some sort of boundary
condition is imposed.  Any particular choice of these conditions will
correspond to a special self-adjoint extension of the Dirac operator
and a special physical input.

We may summarize the above analysis as follows. The $\uparrow$-Dirac
mode in the field of the singular repulsive potential
$+{\cal Q}_{\uparrow}/r^2$ has bound
states of energies $-3/8<E_{\uparrow}{\cal Q}_{\uparrow}<-\infty$.
Furthermore, this field creates its own distribution of pseudoscalar
density ${\cal P}_{\uparrow}$ which tends to amplify the external
potential and to push the field into stronger localization at
{\em negative energies} (the upper bound $E_*$ approaches zero when
${\cal Q}_{\uparrow}$ increases).

Following the same procedure, we may solve the Eqs.~(\ref{eq:E5.1})
and (\ref{eq:E5.3}) for the $\uparrow$-mode
in the field of a $\downarrow$-center,
\begin{eqnarray}
~[E_{\uparrow} -{{\cal Q}_{\downarrow}\over r^2} - i\partial_r] f_{\uparrow}=
m h_{\uparrow},\nonumber\\
~[E_{\uparrow} -{{\cal Q}_{\downarrow}\over r^2} + i\partial_r]h_{\uparrow}=
m f_{\uparrow},
\label{eq:E5.16}\end{eqnarray}
so that now the singular central potential becomes attractive. There
is no need to repeat all the calculations because the change of the sign of
${\cal Q}$ is compensated by the change of the signs of $E$ and $m$.
The difference is that now the bound
states of the upper continuum will tend to compensate the
``external'' charge so that there will be a different number of
bound states. In order to guarantee the positive energy of the vector
fields, the coupling constants of the electron were chosen negative.
If one of the localized states with positive energy is
associated with the electron, then the localized states at
negative energies will interact with the electron as positive
charges. Being close to the original Dirac idea, this picture does
not require a continuous sea of occupied states.
The trends of this oversimplified model clearly point to the existence
of some optimal self-consistent values of $m$ and ${\cal Q}$, which
should be looked for as the eigenvalues of the original system
of equations (\ref{eq:E3.12}) and (\ref{eq:E3.15}).

To this point, the picture of ``falling onto a centre'' renders
the probability of finding a Dirac field at this centre finite.
The distribution of probability density is amazingly
uniform within the range $0<r<r_{max}\sim{\cal Q}\sim 1/m$, so that
a true singularity at the origin is not likely to develop.
However, it is the peak of probability density $ R(0)$, which
determines the magnitude of a gravitational mass and makes it
proportional to the sharpness of the localization of the Dirac field.
The smallest particles are the heaviest, and we indeed have
$ R(0)\sim |E|$. This is evidence
that, even being of a different physical origin, the inertial
($\sim \int T^{00}dV\sim E$) and gravitational
($\sim m\int{\cal P}dV\sim R(0)$) masses of stable particles are
the same, up to a
possible factor which can be absorbed into a ``Newton's gravitational
constant'' $G_N \sim (g^2/M^2)$. Only the full solution of
Eqs.~(\ref{eq:E3.12}) and (\ref{eq:E3.15}) can tell if this factor
is universal. An affirmative answer will be the ultimate proof that
inertial and gravitational masses are equal.
It looks like the mass, which corresponds
to this gravitational constant, is smaller than the formal
``dimensional'' Planck mass, $M_{Planck}^2=(\hbar c/G_N)$, by a factor
$g$, related to the electro-weak interactions. We suggest that the
electro-weak and gravitational constants uniquely determine the mass
$M$ of the axial field.

Finally, it is not clear yet if the localized solutions of the toy model
of this section match by their properties any presently known form of matter,
or even if they are stable. (As it was mentioned in Sec.~\ref{sec:Sec3}, it
is not obvious that the static solutions of sine-Gordon even exist.)
Being strongly localized, such solutions are not likely to effectively
interact with the ``normal'' matter. Nevertheless, these
spherically-symmetric solutions clearly are a source of the gravitational
field and should be considered as the candidates for the dark matter.

\section{ Concluding remarks.}
\renewcommand{\theequation}{6.\arabic{equation}}
\setcounter{equation}{0}

The axial field $\aleph_a$ seems to have been known in connection
with various physical phenomena and under various names for a long
time. The divergence of  this field is proportional to the divergence
of the axial current which carries the quantum numbers of a pion
field. In the non-relativistic limit, the axial field in the Pauli
equation \cite{paper1} adequately describes the parity
nonconservation phenomena in atomic physics. Together with the
Maxwell field in the spin connection it can reproduce the effects of
neutral and charged currents of the standard model provided, as it
was conjectured in Sec.\ref{sec:Sec5}, that the observed sign of the
electric  charge is connected with the sign of energy and
polarization of the localized Dirac states. In other words, the
immediately anticipated physical effects of the axial field seem to
be known or expected.

The new physical {\em quality} is that, like the Maxwell field, the
axial field comes into sight as one of the ingredients of parallel
transport of the  Dirac field. It is a kinematic effect stemming from
the complex nature of the spinor field and a large number of its
polarization degrees of freedom. Truly surprising is the fact that
this field is derived at the most basic level of Lorentz invariance
and that it was not motivated by any phenomenological input, e.g., by
 a need to justify an observed conservation law.

The new result is that the existence of the axial field and an
additional guess that the Dirac field {\em can} form compact objects
(particles) is enough to initiate an effective mechanism of the
auto-localization of the Dirac field.  Indeed, {\em it would be
extremely difficult to find any counterarguments to this guess in a
real world}. If this mechanism is actually realized in Nature, then
all the localized matter in the Universe can be viewed as a
collection of various clusters of the Dirac field, which are tightly
bound by the axial field at the shortest distances and weakly
interact via the same field at the largest ones.  Furthermore, the
existence of a so natural mechanism of auto-localization allows one
to think that the known symmetries and conservation laws can be a
consequence of the soliton nature of the Dirac field.

The nonlinear equations for the Dirac field were advocated long ago
by Heisenberg \cite{Heisenberg} and by Thirring (massless Thirring
model in 1+1 dimension, \cite{Thirring}). The correspondence between
the quantum sine-Gordon equation and the massive Thirring model was
established by Coleman \cite{Coleman}.

The equations of Sec.\ref{sec:Sec3} allow one to derive a
phenomenological Lagrangian of a theory that apparently includes the
{\em axion} field \cite{Weinberg2}. The field $\Upsilon$ and the
axion field couple to the Dirac field in the same way and the
effective Lagrangian includes terms that lead to the sine-Gordon type
equation of motion for the axion field. However, one cannot associate
$\Upsilon$ with a spinless particle because it is a descendent of a
vector field $\aleph_\mu$ in the classical limit.

It was not expected {\em a priori} that in order that the Dirac field
could form compact objects the axial field must be the gravitational
field. However, the picture of gravity as an effect of the axial
field clearly matches all known properties of this interaction. In
addition to equations (\ref{eq:E4.7}) and (\ref{eq:E4.4}) (the
Einstein field equation and the local equivalence of the forces of
gravity and of inertia) this picture explains gravity as a coherent
effect that cannot be screened by any bodies or fields. The
impossibility to screen gravity seems to be a consequence of the
singular nature of the gravitational potential. The ``fall onto a
centre'' is universal and is guaranteed by a special position of the
Newton's force as a potential in the Dirac equation.  Einstein's
field equations, even without the energy momentum tensor of a matter,
correctly match the motion of material bodies as clusters of the
Dirac field. It would be fair to say that the Dirac field is so
natural ``stuffing'' for the Einstein's singularities, that the
previous feeling of the incompleteness of the Einstein-Infeld theory
fades away. Of all possible solutions of the Einstein equations the
physical meaningful are those that have the realistic partners among
the solutions of Dirac equation with the axial field.

An advantage of the new theory is the small number of initial
assumptions and parameters; it exploits only the fact that spinors
realize a representation of a local group of Lorentz transformations.
The hitherto explored consequences of this approach seem to bring us
closer to the explanation of several, yet not perfectly understood
facts:

(i) The Sommerfeld formula of fine structure, which treats the
electron as a point-like charge, nevertheless, reproduces the exact
answer of the Dirac theory. Auto-localization of Dirac field
naturally integrates an image of a classical particle into the full
scope of field theory and demystify this coincidence.

It also eliminates a difficult-to-justify procedure of ``switching
on'' the interaction in the S-matrix version of field theory (see,
e.g. \cite{Bogolyubov})., because the colliding localized particles
indeed begin to strongly interact only after they physically overlap
\cite{SWD}.

The auto-localization totally removes the painful problem of
collinear divergences in radiative corrections, which frustrates
calculations of perturbation theory in the limit of massless charged
particles. Such a limit just does not exist in the real world.

(ii) In Sec.V of paper \cite{paper1} it was pointed out that the
axial field has an effective, with respect to its action on atomic
systems, magnetic equivalent $\vec{B}_{\aleph} =(g/\mu_{B})
\overrightarrow{\aleph}$. If $\overrightarrow{\aleph}$ is the
axial-Newton field of a galaxy, then it may serve as a seed field
from which the effect of the galactic dynamo takes over
\cite{Kulsrud}. Even though this field is very small, there is enough
time to accumulate the effect of magnetic-type polarization or other
parity-odd effects.

The theory probably offers a new insight into such questions as, what
the amount of {\em gravitating matter} in the Universe is, and what
form the matter at the latest stages of stellar evolution and at
gravitational collapse takes.

(iii) V. Gribov \cite{Gribov} advocated the ``falling onto a centre''
phenomenon as a mechanism of quark and color confinement.  At short
distances, the axial field behaves adequately for this hypothesis and
allows one to explore it using much more economic means.

A nearly singular axial potential makes the phase shift between
incoming and outgoing waves of any scattering process (which is not a
hidden parameter) uncertain. One can wonder if this can be a reason
why multiply repeated collisions form a statistical ensemble.
Furthermore, it is not so obvious that equation (\ref{eq:E3.15}) for
the singular part of axial potential has static solutions at all. If
so, then any interaction in real matter is affected by this invisible
from outside temporal dynamics in compact spinor configurations.

\begin{acknowledgments}
I am  grateful to S. Konstein for explaining to me the subtleties of
the issue of self-adjointness and helpful discussions.
\end{acknowledgments}

\end{document}